\documentclass[twocolumn,apjl]{emulateapj}
\pdfoutput=1
\usepackage{amstext}
\usepackage{apjfonts}
\usepackage{amsmath}
\usepackage{amssymb}
\usepackage{graphicx}
\usepackage{ifpdf}

\ifpdf
	\DeclareGraphicsExtensions{.png}
\else
	\DeclareGraphicsExtensions{.eps, .png}
\fi	

\begin{document}

\shorttitle{Halo Retention and Evolution of Coalescing Compact Binaries}

\shortauthors{Zemp, Ramirez-Ruiz \& Diemand}

\title{Halo Retention and Evolution of Coalescing Compact Binaries in Cosmological Simulations of Structure Formation: Implications for Short Gamma-Ray Bursts}

\author{Marcel Zemp\altaffilmark{1,2}, Enrico Ramirez-Ruiz\altaffilmark{2} and J{\"u}rg Diemand\altaffilmark{2,3}}
\altaffiltext{1}{Department of Astronomy, University of Michigan, Ann Arbor, MI 48109; mzemp@umich.edu}
\altaffiltext{2}{Department of Astronomy and Astrophysics, University of California, Santa Cruz, CA 95064; enrico@ucolick.org, diemand@ucolick.org}
\altaffiltext{3}{Hubble Fellow}

\begin{abstract} 
Merging compact binaries are the one source of gravitational radiation so far identified. Because short-period systems which will merge in less than a Hubble time have already been observed as binary pulsars, they are important both as gravitational wave sources for observatories such as LIGO but also as progenitors for short gamma-ray bursts (SGRBs). The fact that these systems must have large systemic velocities implies that by the time they merge, they will be far from their formation site. The locations of merging sites depend sensitively on the gravitational potential of the galaxy host, which until now has been assumed to be static. Here we refine such calculations to incorporate the temporal evolution of the host's gravitational potential as well as that of its nearby neighbors using cosmological simulations of structure formation. This results in merger site distributions that are more diffusively distributed with respect to their putative hosts, with locations extending out to distances of a few Mpc for lighter halos. The degree of mixing between neighboring compact binary populations computed in this way is severely enhanced in environments with a high number density of galaxies. We find that SGRB redshift estimates based solely on the nearest galaxy in projection can be very inaccurate, if progenitor systems inhere large systematic kicks at birth.
\end{abstract}


\section{Introduction}

The association of short gamma-ray bursts with both star-forming galaxies and with ellipticals dominated by old stellar populations \citep{Berger05,Bloom05,Fox05,Gehrels05,Prochaska05,Berger09} suggested an analogy to type Ia supernovae, as it indicated a class of progenitors with a wide distribution of delay times between formation and explosion. Similarly, just as core-collapse supernovae are discovered almost exclusively in star-forming galaxies, so too are long GRBs \citep{wbrev}. Indeed, a detailed census of the types of host galaxies, burst locations and redshifts could help decide between the various SGRB progenitor alternatives \citep[e.g.][]{zr07,guetta05,bp06}: If the progenitor lifetime is long and the systemic kick is small, then the bursts should correspond spatially to the oldest populations in a given host galaxy. For early-type galaxies, the distribution would most likely follow the light of the host. A neutron star (NS) binary could take billions of years to spiral together, and could by then, if given a substantial kick velocity on formation, merge far from its birth site \citep{fryer99,Bloom99,bbk02,Bloom02}.  The burst offsets would then most likely be larger for smaller mass hosts.

Double neutron star binaries, such as the famous PSR1913+16, will eventually coalesce, when gravitational radiation drives them together \citep{vk07}. Each supernova is thought to impart a substantial kick to the resulting NS \citep{hp97}.  For systems that survive both supernovae explosions the center of mass of the remnant binary itself will receive a velocity boost on the order of a few hundred kilometers per second \citep{bp95,fk97}.  As a result, NS binaries will be ejected from their birth sites.  The exact distribution of merger sites depends sensitively on the gravitational potential of the host, which until now has been assumed to be static. The potential of a realistic host galaxy is, however, not static. In fact, the gravitational potential of the host as well as that of its nearby neighbors is expected to evolve dramatically from compact binary production until coalescence.  In order to incorporate these effects self-consistently, in this {\it Letter}, we study the orbital evolution of compact binary systems using cosmological simulations of structure formation.  Our results provide new insights into what happens when compact binary stars are ejected from their birth halos as a result of velocity kicks, and what progenitor clues a distant observer might uncover from the distribution of SGRB sites in and around galaxies.

\section{Compact Binaries in Cosmological Simulations}

The focus of this work is to understand the retention and evolution of compact binaries in an evolving cosmological simulation. To this end, we have performed a dark matter only cosmological structure formation simulation. A 80 comoving Mpc periodic box with a single mass resolution of $m_p = 1.07 \times 10^{9}~M_\odot$ corresponding to 256$^3$ particles is initialized at a starting redshift of z = 22.4 (161 Myr). The particles have a softening length of 16 kpc and we have used the WMAP 3-year cosmological parameters \citep{2007ApJS..170..377S} with $\Omega_{\mathrm{M},0} = 0.238$, $\Omega_{\Lambda,0} = 0.762$ and $H_0 = 73~{\rm km}~{\rm s}^{-1}~{\rm  Mpc}^{-1}$, $\sigma_8 = 0.74$ and $n_s = 0.951$. The time evolution was done with the parallel tree-code PKDGRAV2 \citep{2001PhDT........21S}. PKDGRAV2 uses a fast multipole expansion technique in order to calculate the forces with a hexadecapole precision and a time-stepping scheme that is based on the true dynamical time of the particles with an accuracy parameter of $\eta_{\mathrm{D}} = 0.03$ \citep{2007MNRAS.376..273Z}.

First, we evolve the initial conditions until redshift z = 1.60 (4.24 Gyr). At this time we find all the halos in our simulation with a friends-of-friends method \citep{davis} and select all those with a minimum of 200 particles of which there are 2461. This criterion corresponds to a halo mass of $2.15 \times 10^{11}~M_\odot$. Second, we populate each of these selected halos with 2000 massless tracer particles which are placed at the centre of each halo. Each tracer particle is meant to represent a compact binary system, which, on average, forms around the peak of the star formation epoch \citep{Madau96}.  The velocity distribution of the tracer particles is assumed to be isotropic and to have a Maxwell-Boltzmann distribution with a mean speed of $\bar{v} = 360~{\rm km}~{\rm s}^{-1}$ and a dispersion of $\sigma = 150~{\rm  km}~{\rm s}^{-1}$. This is consistent with the magnitude of the natal kicks required to explain the observed parameters of binary neutron star systems -- only when kicks have magnitudes exceeding 200 km s$^{-1}$ can the progenitor orbits be sufficiently wide to accommodate evolved helium stars and still produce the small separations measured in these systems \citep{bp95,fk97}. Third, we weight each tracer particle by $w_i \equiv m_i / m_{\rm max}$ where $m_i$ is the mass of the halo where the tracer particle with index $i$ initially starts and $m_{\rm max}$ is the mass of the most massive halo at redshift z = 1.60.  This is done in order to account for the contribution of a given halo to the total number of compact binaries produced at z = 1.60 under the assumption that it scales with the mass content in each halo. Finally, we evolve the cosmological cube together with the tracer particle populations until redshift z = 0 (13.8 Gyr).

To explore the ability of individual halos in retaining their own birth compact binary population, one requires to accurately follow not only the location of the tracer particles but also the fate of the individual halos as they evolve and experienced a substantial metamorphosis. To this end, we have marked the particles with the highest phase space density in each of the 2461 halos that have been populated with tracer particles. By following these marked particles, we can then accurately track the location of the halos over time. These high phase space density particles are optimal for tracking the halo positions since they stay at the centre of the
halo and are only minimally affected by tidal effects. Even when a halo becomes tidally disrupted, the centre of the debris of these high phase space density particles still provides a good representation of where the halo would be if it would not have been disrupted.

\section{The Cosmological Compact Binary Merging Sites}

\begin{figure}
	\centering
	\includegraphics[width=\textwidth]{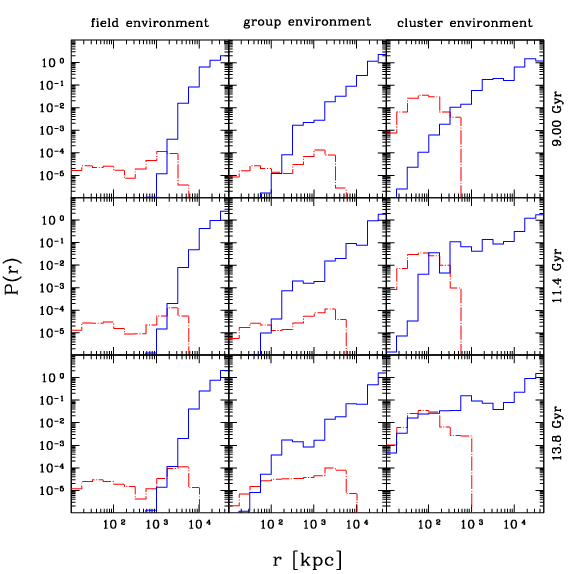}
	\caption{Probability density function $P(r)$ of the tracer particles as a function of distance from a particular halo for three different environments and at three different snapshots in time. The dash-dotted red lines show the radial probability density function for all tracer particles formed in the particular halo while the solid blue lines show the same function for tracers belonging to all other halos. Note how the bound and unbound tracers of local origin (dash-dot/red) give rise to bimodal distributions in the field and group environments, while all of them remain bound in cluster environment.}
	\label{fig1}
\end{figure}

We now turn to an examination of the cosmological distribution of the compact binary systems, which is done here by following the massless tracer particle populations over time. Herefore, we picked out by eye three halos, each in a different environment (field, group and cluster). Figure \ref{fig1} shows the probability density function of finding tracer particles as a function of distance from a given halo at three different snapshots in time. The red lines show the radial probability density function for all tracer particles originating in the particular halo while the blue lines show the same function for tracers belonging to all other halos.

Figure \ref{fig1} is self-explanatory. In the field environment, the distribution is dominated out to a distance of a few Mpc by its own tracer particles at all cosmological times. The tracer particle population injected at z = 1.60 (4.24 Gyr) is observed to be rather extended $\sim 10~{\rm Mpc}$. This is because this field halo, whose mass at z = 1.60 was $2.15 \times 10^{11} M_\odot$, was unable to effectively retain most of its own tracer particles. This is also the case for the host halo belonging to the group environment, whose central mass at z = 1.60 was $2.27 \times 10^{11} M_\odot$. However, in this more crowded environment, the close proximity of the neighboring halos allows foreign tracer particles to pollute the central regions of the host halo.  In the cluster environment, only very few tracer particles are able to escape the deep halo potential well where they were born \citep{niino}, whose mass at z = 1.60 was $5.95 \times 10^{13} M_\odot$. The few unbound particles are still effectively retained by the cluster's potential. As a result of the high merger activity in such cluster environments, the mixing of the various tracer populations increases dramatically with time. At $z=0$, for example, the probability of finding a foreign tracer is equal or higher at all radial distances than finding one originating from the massive central halo, whose mass at z = 0 is now $7.64 \times 10^{14} M_\odot$. This clearly indicates, that in a high density environment with a lot of merger activity there is a high degree of tracer mixing and, as a result, the closest galaxy at the time of merging is likely not to be the one where the compact binary system originated from.

\begin{figure}
	\centering
	\includegraphics[width=\textwidth]{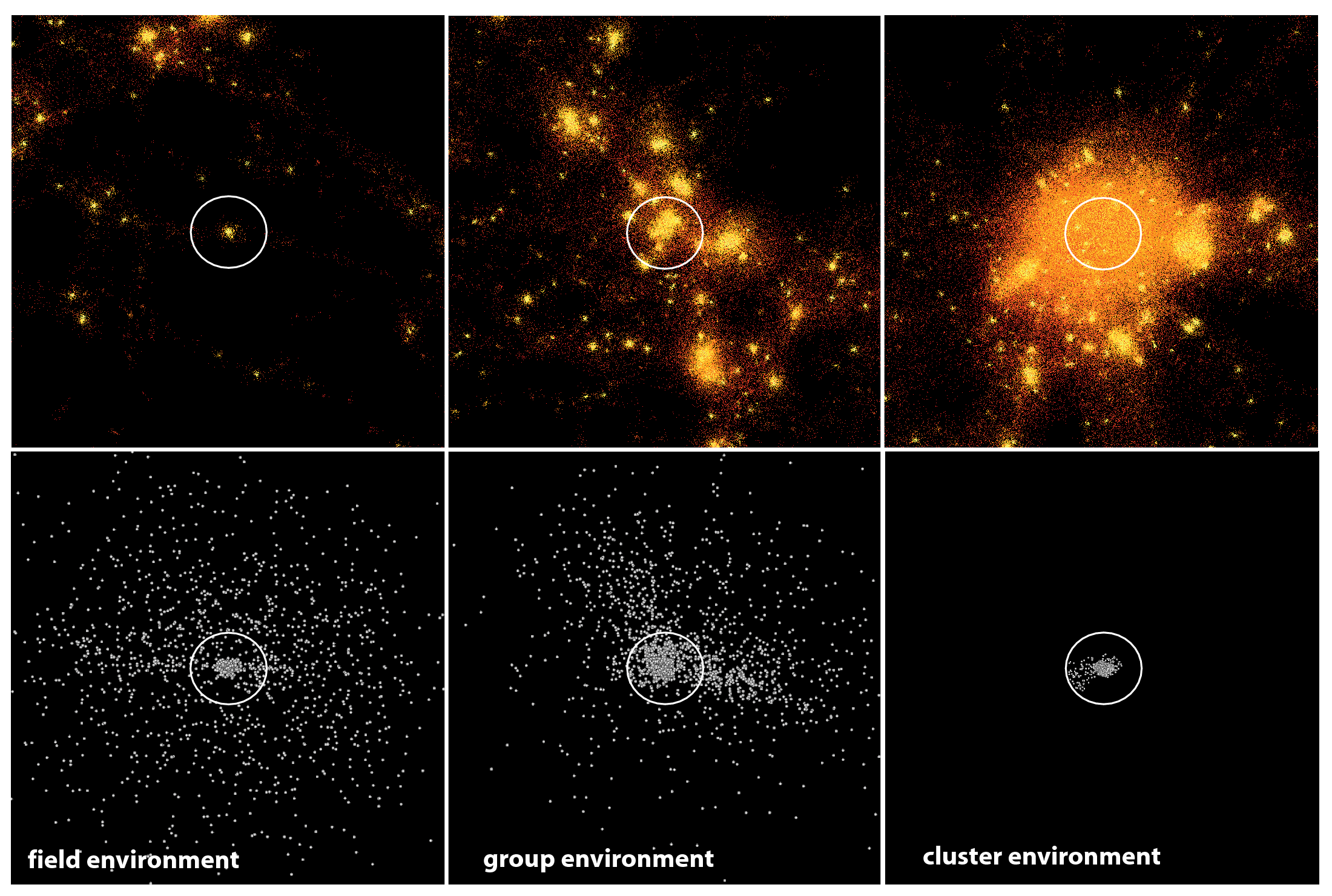}
	\caption{The tracer particle populations as a function of environment in the local universe. The dark matter density field at z = 0 (top row) of the three halo environments shown in Figure \ref{fig1} are plotted together with the distribution of the particle tracers that originate from the selected central halo (bottom row). The  side length of each panel is 10 Mpc.}
	\label{fig2}
\end{figure}

The ability of a halo to retain its birth population of tracer particles is further illustrated in Figure \ref{fig2} where we plot the density field at z = 0 (left column) of the three halo environments shown in Figure \ref{fig1} together with the distribution of the particle tracers that originate from the selected halo (right column). We see that in the cluster environment, the tracer particle population stays compact. In this crowded environment, the many accreted subhaloes contribute significantly to the pollution of the tracer particle population in the central cluster region as shown in Figure \ref{fig1}. In the small mass field halo, the tracer particles spread over many Mpc in distance. Given that there are not many other halos in the immediate neighbourhood, the tracer particle pollution is negligible. This is, however, not the case for the selected halo in the group environment which despite having a mass similar to that of the field halo at z = 1.60, experiences a substantial degree of tracer particle mixing from the other group members.

\begin{figure}
	\centering
	\includegraphics[width=\textwidth]{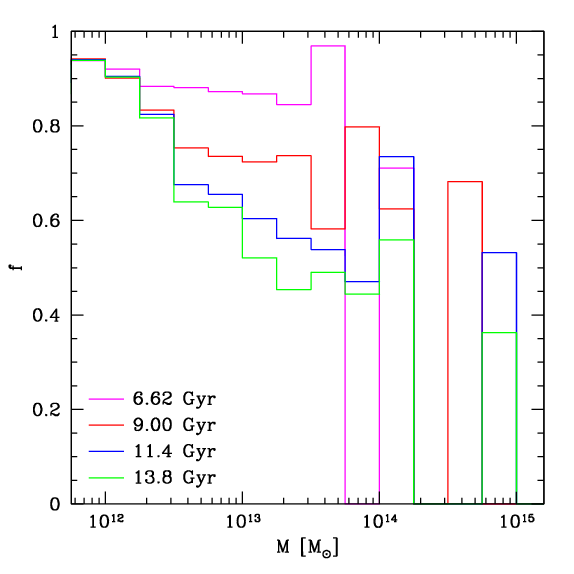}
	\caption{The weighted fraction $f$ of tracer particles over time that are still closest to the halo they started at redshift z = 1.60 as a  function of the mass of the most massive nearby halo (see text for details).}
	\label{fig3}
\end{figure}

Now the question arises over whether the compact binary was born in the galaxy it is closest to at the time of merger? If SGRB are triggered by compact binary formation, then the merging site can be deduced by the afterglow location. We can answer this by following the population of compact binaries as the cosmological simulation evolves. At any given time we can calculate the weighted fraction of tracer particles that are still closest to the halo they started at redshift z = 1.60 given by $f \equiv {\sum_k^{N_{\mathrm{co}}} w_k} / {\sum_i^{N_{\mathrm{tot}}} w_i}$, where $N_{\mathrm{tot}} = 2461 \times 2000 = 4922000$ is the total number of tracer particles and the $k$ summation runs over all $N_{\mathrm{co}}$ tracers that are still closest to their original halo. In Figure \ref{fig3}, we plot this fraction as a function of the environment mass containing the tracer particle population.

The environment of a tracer particle is defined by the most massive halo we find at a given time that contains the tracer particle. We define that a tracer particle is contained within a halo if they are within a sphere of radius $d=2s$ from its centre and where $s$ is given by $s \equiv \frac{1}{6} \left(\sum_{k=1}^3 x_{k,\mathrm{max}} - x_{k,\mathrm{min}}\right)$ and $x_{k,\mathrm{max}}$ respectively $x_{k,\mathrm{min}}$ are the maximum/minimum coordinate in the $k$th dimension of any particle in the halo. The fraction $f$ is not very sensitive to the detailed definition of environment used here since for example the definition $d=s$ results in a qualitatively similar plot.

Figure \ref{fig3} highlights the importance of environment on compact binary retention. It shows that in environments $\lesssim 10^{12}~M_\odot$ around 95\% of all weighted tracers at all cosmological times are still closest to their halo where they originate from. For more massive environments, this ceases to be true. For example, in environments $\gtrsim 10^{14}~M_\odot$, the fraction of binaries found closest to their birth halo severely decreases with time, reaching 45\% at redshift z = 0.  This implies that in more than half the cases the merger event would not take place closest to its birth galaxy site if occurring within a cluster environment. 

\section{Discussion}

Before the detailed modeling of light curves was used to constrain the nature of supernovae progenitors, the location of supernovae in and around galaxies provided important clues to the nature of the progenitors.  Similarly, in the absence of supernova-like features \citep[e.g.][]{hjorth05a}, detailed observations of the astrophysics of individual host galaxies may thus be essential before stringent constraints on the identity of SGRB progenitors can be placed\footnote{The interaction of burst ejecta with a stellar binary companion \citep{andrew} or with its emitted radiation \citep{enrico} could also help shedding light on the identity of the progenitor system.} \citep[e.g.][]{zr07,Fox05}. 

Even with a handful of SGRBs detected to date, it has become apparent that short and long events are not drawn from the same parent stellar population \citep{nakar}. In contrast to long GRBs, the galaxies associated with SGRBs exhibit a wide range of star-formation rates, morphologies and metallicities \citep{Berger09}. They are also often found in older and lower-redshift galaxies and, in a few cases, with large ($\gtrsim$ 10 kpc) projected offsets from the centers of their putative host galaxies \citep{Bloom05,Prochaska05,Berger05,Gorosabel06,levan,bloom07}.  SGRBs are, however, not universally at large offsets and are not always associated with early-type galaxies \citep{Fox05,villa,hjorth05b,covino,as06}.

The discovery of early-type galaxy hosts suggest a progenitor lifetime distribution extending well beyond a Gyr. A large progenitor lifetime would help explain the apparent high incidence of galaxy cluster membership \citep{Bloom05,Ped05,Berger07a}. On the other hand, shorter lifetimes are required to explain the population of SGRB at moderately high redshift \citep{Berger07b,graham09}.  The observed projected distances from what has been argued are the plausible hosts, if true, also holds important ramifications for the sort of viable progenitors \citep{bp06}.  The large offsets seen from early-type hosts would seem to be at odds with progenitor systems with small systematic kicks \citep[such as in globular clusters][]{Grindlay06,lrr07,rion}, although with such large physical offsets the possibility remains that the association with the putative host is coincidental. On the other hand, based on the small offsets from some low-mass galaxy hosts \citep{Prochaska05,as06,Berger09,bp06}, SGRB progenitors cannot all have large systematic kicks at birth and inhere large delay times from formation.

Making more quantitative statements about the nature of the progenitor systems is not only hampered by small number statistics but also from the lack of robust predictions of the distribution of merger sites. These distributions depend sensitively on the gravitational potential of the host, which until now has been assumed to be static, and the compact binary formation properties, especially the systematic kick velocity. 

Here we have refined such calculations to incorporate the temporal evolution of the host's gravitational potential as well as that of its nearby neighbors self-consistently using cosmological simulations of structure formation. This results in diverse predictions of offsets and compact binary demographics even in the simplest case of a kick velocity distribution \citep[here assumed to be in excess of $200\;{\rm km\;s^{-1}}$ in order to explain the observed parameters of double neutron star systems;][]{fk97} whose properties do not vary with the initial binary separation. 

Two important predictions stand out. First, the merger site distributions computed in this way are more diffusively located with respect to their putative hosts. In a field environment, for example, the distribution of merging sites can extend out to a distance of a few Mpc. This is more severe for those host galaxy halos that were unable to effectively retain most of its own compact binary population at birth. Second, the degree of mixing between neighboring compact binary populations depends on galactic environment. In a cluster, for example, the mixing of the various compact binary populations is severe as a result of the high merger activity and increases dramatically with time. At $z=0$, in a cluster environment, the probability of finding a foreign coalescing compact binary system is equal or higher at all radial distances than finding one originating from the massive central halo. As a result, the closest galaxy at the time of binary coalescence and possibly SGRB occurrence may not to be the one where the compact binary system originated from.

Of course, our basic model is rather simple since we assume a single epoch of star formation and a simple star formation recipe. Also different distributions of kick velocities should be considered. We do not expect that our qualitative results will change dramatically as our first, more general, results indicate.

It is evident form the discussion above that assuming a large (already evolved) host galaxy at the time of compact binary formation thus severely overestimates the binary retention fraction and the concentration of their merging site distribution. This implies that SGRB redshift estimates based solely on the nearest galaxy in projection can be very inaccurate, if progenitor systems inhere large systematic kicks at birth. Interpretations on the nature of the SGRB progenitor using the stellar and mass properties of the nearest galaxy in projection as established by the afterglow location must therefore be regarded with suspicion.  Finally, it should be noted that a direct comparison with model predictions is still impeded by the possibility of an ambient density bias \citep{Bloom05,Lee05} where SGRBs are more likely to be found in denser gas regions and , as a result, we could be missing a population of bursts with large systematic kicks.

\acknowledgments We thank J. Bloom, S. Faber, C. Fryer, V.Kalogera, R. O'Shaughnessy and X. Prochaska for useful discussions.  M.Z. is supported by NSF grant AST-0708087. We further acknowledge support from Swift: NASA NNX08AN88G (E.R.), NSF: 0521566, and the David and Lucile Packard Foundation (E.R.).

\newpage

\newpage

\newpage

\end{document}